\documentclass[pre,twocolumn,amsmath,amssymb]{revtex4-1}

\usepackage{graphicx} 
\usepackage{enumitem}
\usepackage{float}
\usepackage{physics}
\usepackage{amsmath}
\usepackage{balance}
\usepackage{color}
\usepackage{lastpage}
\newcommand{\red}{\textcolor{black}}
\newcommand{\redd}{\textcolor{black}}

\usepackage[colorlinks=True]{hyperref}  
\hypersetup{
 	colorlinks=true,  
 	linkcolor=blue,  
 	citecolor=blue, 
 	filecolor=magenta,   
 	urlcolor=blue         
 }
 
\makeatletter
\def\maketitle{
\@author@finish
\title@column\titleblock@produce
\suppressfloats[t]}
\makeatother

\makeatletter
\newcommand\footnoteref[1]{\protected@xdef\@thefnmark{\ref{#1}}\@footnotemark}
\makeatother

\makeatletter
    \def\balanceissued{unbalanced}
    \let\oldbibitem\bibitem
    \def\bibitem{%
        \ifnum\thepage=6%
            \expandafter\ifx\expandafter\relax\balanceissued\relax\else%
                \balance%
                \gdef\balanceissued{\relax}\fi%
            \else\fi%
        \oldbibitem}
\makeatother

\begin{document}
\title{\red{Specialization} at an expanding front} 

\author{Lauren H. Li} 
\author{Mehran Kardar}
\affiliation{Department of Physics, Massachusetts Institute of Technology, Cambridge, Massachusetts 02139, USA}
\date{\today}
\date{September 13, 2023}


\begin{abstract}
    As a population grows, \red{spreading to new environments may favor specialization.} In this paper, we introduce and explore a model for \red{specialization} at the front of a colony expanding synchronously into new territory. We show through numerical simulations that, by gaining fitness through accumulating mutations, progeny of the initial seed population can \red{differentiate into distinct specialists.} With competition and selection limited to the growth front, the emerging \red{specialists} first segregate into sectors, which then expand to dominate the entire population. We quantify the scaling of the fixation time with the size of the population and observe different behaviors corresponding to distinct universality classes: unbounded and bounded gains in fitness lead to superdiffusive ($z=3/2$) and diffusive ($z=2$) stochastic wanderings of the sector boundaries, respectively.
\end{abstract}

\maketitle
\red{In the course of evolution, a homogenous population may diversify to exploit emerging ecological niches. Such disruption of a population's homogeneity can often be attributed to changes in the availability of resources across geographical terrains~\cite{hall2007does}. As an initially homogeneous population occupies new terrain, it can differentiate into different specialized populations to maximize fitness.} Here, we introduce a fitness model for specialization by mutations along a ``two-feature" axis. In our model, the mutating population expands spatially at a front (similar to a tumor), with reproductive selection encoded by a fitness function. Positing different forms of the fitness function, we use numerical simulations to follow the evolution of the population, in particular tracking the fixation time for the entire population to become dominated by a single \red{specialized group.} Our main result is that, depending on whether the fitness can grow indefinitely or there exists a maximum attainable fitness, the fluctuation behavior of domain walls between \red{specialized populations} falls into different universality classes. 

Our work is largely inspired by studies of \textit{range expansions}, which describe populations that expand spatially into new territory over the course of many generations, as in tumors~\red{\cite{bru2003universal, haeno2010evolution}} or bacterial colonies~\red{\cite{hallatschek2007genetic, lee2022slow}.} In these toy models of expanding populations, reproducing individuals only compete with those in close proximity at the front of the expanding colony, and the effects of genetic drift are amplified due to this spatial limitation~\cite{korolev2010genetic}. As such, range expansions provide rich arenas for studying stochasticity in evolution and have motivated numerous works~\cite{korolev2011quantitative, drossel2000phase, hallatschek2010life, korolev2012selective, farrell2017mechanical, horowitz2019bacterial, hallatschek2007genetic, saito1995critical, korolev2010genetic, gralka2016allele, chu2019evolution}. In laboratory experiments, Hallatschek \textit{et al}.~\cite{hallatschek2007genetic} studied the appearance of sectors in growing bacterial populations: an initially well-mixed population of two fluorescently labeled strains of \textit{Escherichia coli} \redd{was} allowed to grow and expand. After some time, cells at the colonization front had segregated into sectors defined by the fluorescent marker. Interestingly, the mean square transverse displacements of the sector boundaries scaled with expansion radius $\ell$ as $\ell^{2\zeta}$ with exponent $\zeta=0.65\pm0.05$, suggesting \textit{superdiffusive} wandering with $\zeta=2/3$ in the Kardar-Parisi-Zhang (KPZ) universality class~\cite{kardar1987replica}. Furthermore, such observations are not unique to \textit{E.~coli}, as similar experiments with growing colonies of haploid \textit{Saccharomyces cerevisiae} revealed the same scaling~\cite{hallatschek2007genetic}. Hence, superdiffusive behavior is hypothesized to be a universal characteristic of \red{certain} microbial range expansions \red{\footnote{\red{While noting that Refs.~\cite{korolev2010genetic} and \cite{korolev2011quantitative} report different scaling exponents, this work concerns experimental and simulation studies of bacterial colonies that exhibit superdiffusive behavior.}}.}

Several numerical studies on simple models of growth have elucidated the universal characteristics of range expansions. One well-studied class of models are stepping stone models~\cite{kimura1964stepping}, which represent the growing colony with occupied points on a lattice, with sites at the front reproducing into neighboring unoccupied sites~\cite{saito1995critical, korolev2010genetic, gralka2016allele, chu2019evolution}. With layer-by-layer (synchronous) growth (starting from a straight one-dimensional edge), the boundary between two sectors performs a random walk corresponding to a transverse roughness exponent of $\zeta=1/2$; however, asynchronous growth (random selection of sites on the front) results in a rough front and leads to the superdiffusive exponent $\zeta=2/3$~\redd{\cite{chu2019evolution}}, as is the case in experiments~\redd{\cite{hallatschek2007genetic}}. Moreover, the superdiffusive exponent is also observed in a model with synchronous reproduction~\cite{chu2019evolution} inspired by directed paths in random media (DPRM)~\cite{kardar1987scaling}. The latter model can be interpreted as describing stochastic variations in the size of the cells, giving rise to a rough front \red{\footnote{\red{DPRM and rough interface growth are two prominent examples of the \textit{KPZ universality class}. While there are a number of signatures, an important characteristic of this class is super-diffusive ($\delta x\sim t^{2/3}$) as opposed to diffusive ($\delta x\sim t^{1/2}$) spread of fluctuations}}.} As a variant of the latter model, our work helps clarify when synchronous reproduction can result in superdiffusive or diffusive scaling.  

In the above range expansion experiments, the different sectors can be regarded as distinct \red{specialized populations} competing at the sector boundaries. However, all these \red{specialists} are already present at the initial seed (with assumed identical fitness) and their progeny then segregate into different regions of space. In contrast, \red{specialization} in nature typically arises from the \red{differentiation of individuals over time.} In this letter, we introduce a model in which distinct \red{specialists} evolve spontaneously due to mutations; subsequent competition with neighboring \red{specialists populations} ensues and leads to spatial separation. Our model is consistent with observations in nature and in experiments. 

\emph{Fitness model.} We consider a simple model with individuals characterized by two traits, which we label as breadth $b$ or height $h$; the reproductive success of individuals in competition is given by some function $f[b,h]$ of the two traits. Our focus is on whether mutations lead to \red{individuals specializing} in one trait over the other, and hence we are interested in the difference in magnitude of the two traits $m=b-h$. \red{Studies of phase transitions indicate that key universal features can be captured by simple polynomial expressions for the free energy density~\cite{landau2013statistical}. Motivated by such, we consider a simple polynomial form for the fitness function that allows for differentiation:}
\begin{equation} \label{eq:fit}
    f[b,h] \equiv f(m)= \red{f_0} + \alpha m^2 + \beta m^4 + \gamma m.
\end{equation}
The even powered terms ($\alpha$ and $\beta$) preserve a symmetry about $f_0$ between $b$ and $h$ (or $m\to-m$), while the odd term ($\gamma$) breaks this symmetry \red{\footnote{\red{We set $f_0=0$ without loss of generality.}}.} The degree of specialization is quantified by $m$, which is akin to the magnetization in Landau's theory of magnetic phase transitions~\cite{landau2013statistical}. 

\emph{Lattice implementation.} Given the fitness function in Eq.~\eqref{eq:fit}, we model spatial growth using a variation of the stepping stone model~\cite{kimura1964stepping}, where individuals in generation $t$ are arranged along a line, indicated by $x$; we refer to individuals by their site coordinates $(x,t)$ on a triangular lattice (Fig.~\ref{fig:update}). The progeny at generation $t$ are determined by the competition between the two neighboring individuals at generation ($t-1$). The winner of the competition between potential parents ($x+1,t-1$) and ($x-1,t-1$) is the one with the larger fitness value
\begin{equation}
    f(x,t) = f[b(x,t), h(x,t)].
\end{equation}	
Let $x_{max}$ denote the $x$ coordinate of the individual with the higher $f(x,t)$. The individual at $x_{max}$ then procreates and its progeny inherits its traits, up to small variations as described by
\begin{equation}
    \begin{aligned}
        b(x,t) &= b(x_{max}, t-1) + \eta_b(x,t)\,,\\
        h(x,t) &= h(x_{max}, t-1) + \eta_h(x,t)\,.
    \end{aligned}
\end{equation}
We take $\eta_b(x,t)$, $\eta_h(x,t)$ to be independent and identically distributed Gaussian random variables, with zero mean and correlations $\langle\eta_a(x,t)\eta_{a'}(x',t')\rangle =\sigma^2\delta_{a,{a'}} \delta_{x,{x'}} \delta_{t,{t'}} $, as in the DPRM-inspired model in Ref.~\cite{chu2019evolution}. The noise $\eta$ accounts for random mutations in the offspring. Since the mean change is zero, the magnitudes of $b$ and $h$ are equally likely to increase or decrease over generations; however, the accumulating mutations selected by preferential reproduction enable specialization to occur in certain regimes of the fitness function. Note that as all individuals in a given generation $t$ are updated synchronously, the uppermost layer in Fig.~\ref{fig:update} remains flat. 

\begin{figure}
    \centering
    \includegraphics{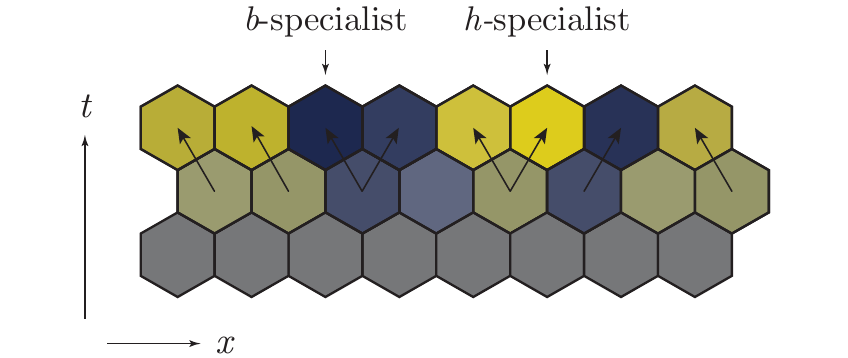}
    \caption{Illustration of the update rules for our model. The initial seed population ($t=0$) is unspecialized (grey), and subsequent progeny inherit features $b$ and $h$, according to our update rules. Variations provided by the accumulating random mutations in the update rules may result in individuals becoming specialized in $b$ (blue) or $h$ (yellow).}
    \label{fig:update}
\end{figure}

Initially, all members of the population are unspecialized; that is, $m(x,0)=0$ for all $x$. For some fitness functions, individuals may become specialized in either feature, such that after $t$ generations, $m(x,t)\neq0$ for typical $x$ (Fig.~\ref{fig:update}). Over time, segments of neighboring individuals with the same specialization form sectors.
Depending on the shape of the fitness function (parametrized by $\alpha$, $\beta$, and $\gamma$) and the magnitude of mutations (parametrized by $\sigma$), we observe different patterns in the emergence of new specialist populations (Fig.~\ref{fig:regimes}).

\begin{figure*}
    \centering
    \includegraphics{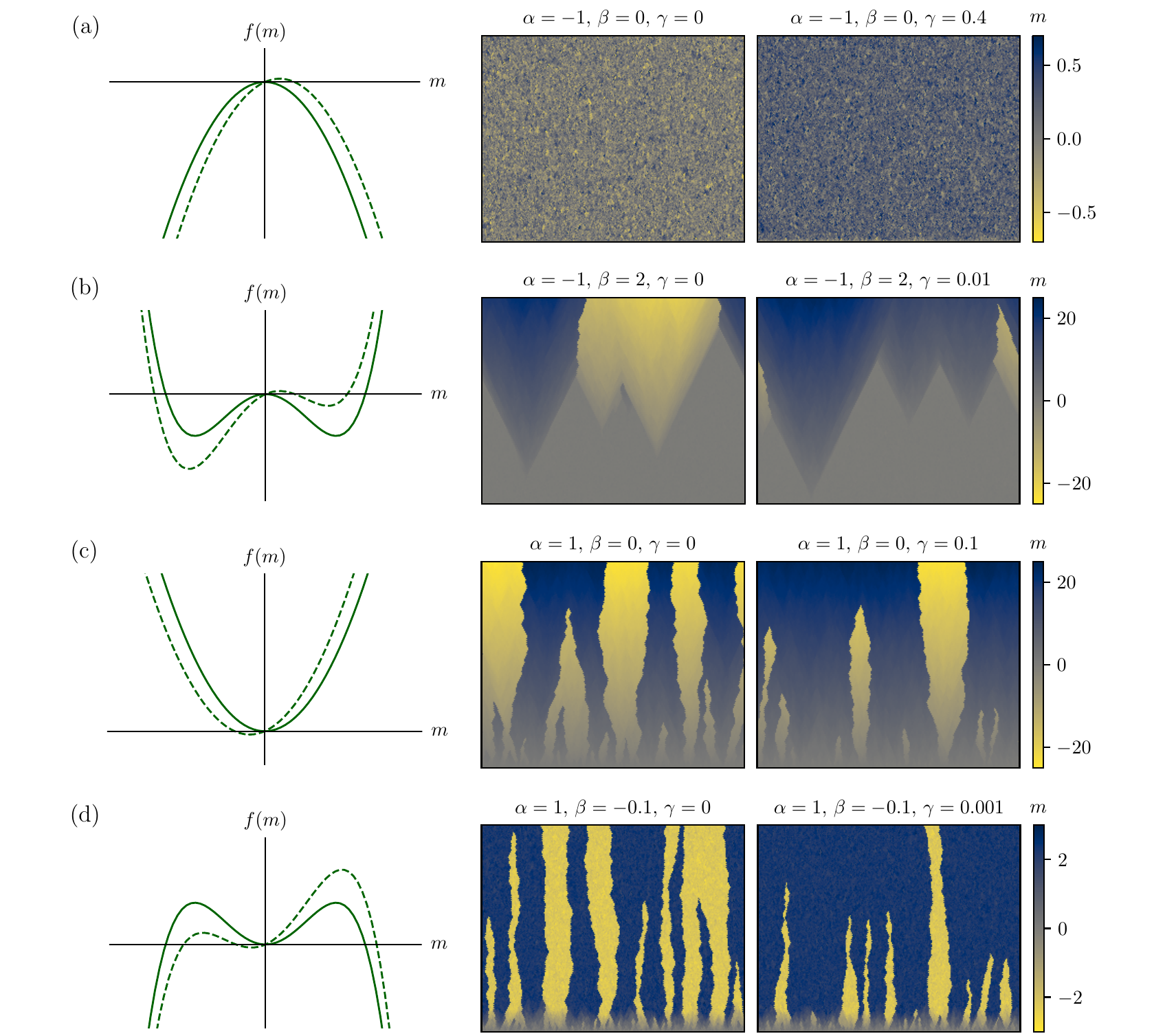}
    \caption{Possible forms of the fitness function $f(m)$ without ($\gamma=0$, solid) or with ($\gamma>0$, dashed) symmetry breaking.  The fitness function favors (a) generalists; (b) specialists, although an initial advantage exists for generalists; (c) specialists, with unbounded fitness gain; and (d) specialists, with bounded fitness gain. For each of the four cases, we observe the corresponding evolutionary dynamics generated by our update rules (with periodic boundary conditions along the horizontal direction) illustrating different specialization patterns.  Color plots show $m(x,t)$ for populations of size $L=256$ over $200$ generations, with time running in the upward direction. We set the standard deviation of the mutational variability $\eta$ to $\sigma=0.1$ in all plots.}
    \label{fig:regimes}
\end{figure*}

\textit{Growth Patterns.}  We conduct numerical simulations with our fitness model under different regimes. In Fig.~\ref{fig:regimes}(a), the fitness is maximized at the origin, and we observe no \red{specialization.} Even when the fitness is shifted to favor a particular feature by setting $\gamma\neq0$, there is no \red{differentiation into distinct specialized populations; rather,} the population is dominated by specialists in that feature, but these specialists are evenly distributed in space. \red{These behaviors are reminiscent of antibiotic resistance, which can appear or disappear in cells depending on the presence of antibiotics in the medium~\cite{chung2022rapid}.}

In contrast, in Fig.~\ref{fig:regimes}(b), there is a local maximum around the origin with local minima on both sides; hence, it is possible to attain a higher fitness by moving through a less favorable region in the fitness landscape. Due to this partial advantage for generalists, especially at early times, the resulting growth pattern consists of unspecialized individuals until the sudden onset of those highly specialized in $b$ or $h$. Eventually, these specialists form V-shaped sectors which, upon meeting, compete for dominance \red{\footnote{We note that such V-shaped growth patterns are common; a recent study of \textit{Raoultella planticola} also observed such patterns corresponding to the emergence of a more fit mutant strain in a wild-type population~\cite{lee2022slow}.}.}

In Figs.~\ref{fig:regimes}(c) and (d), we observe specialization in cases where the fitness gain is unbounded or bounded, respectively. By random chance, $b$ and $h$ fluctuate over generations, resulting in individuals specializing in one feature over the other, giving rise to \red{specialists.} For the bounded fitness function illustrated in Fig.~\ref{fig:regimes}(d) with two local maxima, we observe an emergence time $\tau_e$ for the magnitude of specialization $\abs{m}$ to become maximal (detailed in Ref.~\footnote{See Supplemental Material for a qualitative analysis of the emergence time and higher-order cumulants of the fixation time.}).

\textit{Fixation time.} Once \red{specialized groups} are well-defined, the domain walls between distinct groups fluctuate as the population expands. Eventually, one \red{specialized population} dominates, and there is a fixation time $\tau_f$ in which the entire population becomes specialized in the same feature~\footnote{The fitness function in Fig.~\ref{fig:regimes}d is reminiscent of the (negative) free energy of a magnet. Since there is no strict phase transition in an equilibrium magnet in one-dimension, we may well inquire if the non-equilibrium model presented here leads to true specialization with no reversion in the $L\to\infty$ limit. Since the corresponding length scales are beyond the range of our simulations (and scope), we did not explore this question.}. We perform numerical simulations to characterize the scaling of the fixation time with population size $L$ as
\begin{equation}
    \tau_f\propto L^z\,,
\end{equation}
with a dynamical exponent $z=1/\zeta$. We find different values for $z$ corresponding to different universality classes depending on the shape of the fitness function.

\begin{figure}
    \centering
    \includegraphics{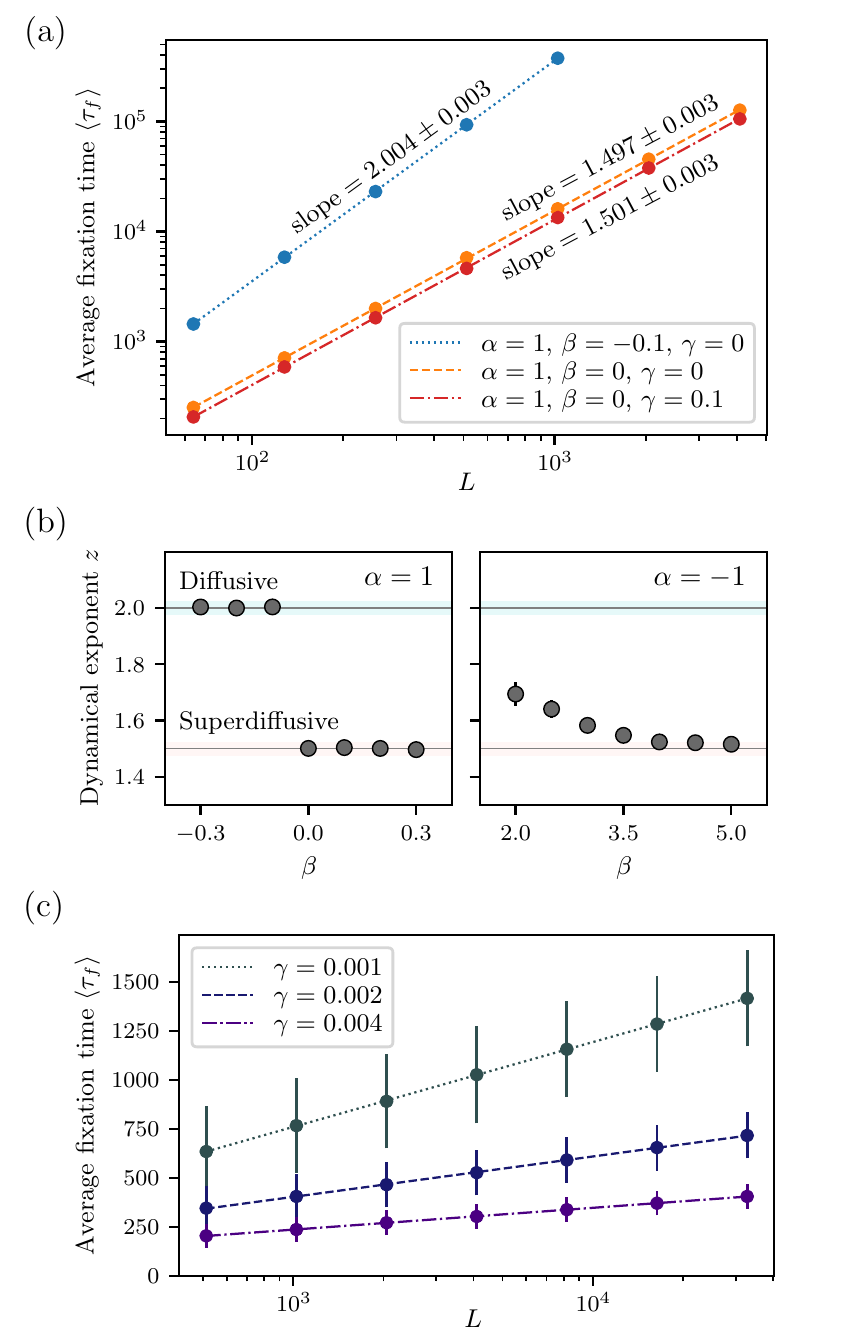}
    \caption{Distinct scaling behaviors (universality classes) of the fixation time. Linear fits indicate the dynamical exponent $z$. (a) Unbounded fitness gain without ($\gamma=0$, dashed) or with ($\gamma=0.1$, dot-dashed) symmetry breaking for $\alpha=1$, $\beta=0$ show superdiffusive scaling with $z\approx3/2$; bounded symmetric fitness gain ($\gamma=0$, dotted) for $\alpha=1$, $\beta=-0.1$ exhibits diffusive behavior with $z\approx2$
    (b) In the symmetric case ($\gamma=0$), the transition between diffusive and superdiffusive regimes is abrupt on changing $\beta$ from negative to positive for $\alpha=1$; a gradual crossover is observed for increasingly positive $\beta$ for $\alpha=-1$. (c) Symmetry breaking ($\gamma\neq0$) in the case of bounded fitness gain leads to rapid fixation time logarithmic in size, with slope inversely proportional to $\gamma$; here, we set $\alpha=1$, $\beta=-0.1$. \redd{Error bars show the standard deviation $\sqrt{\langle\tau_f^2\rangle_c}$.} Statistics are calculated over $10^4$ realizations, with $\sigma=0.1$ in all plots.} 
    \label{fig:scaling}
\end{figure}

In the case of unbounded specialization [Fig.~\ref{fig:regimes}(c)] and $\gamma=0$, the average fixation time $\langle\tau_f\rangle$ scales in $L$ with exponent $z=1.497\pm0.003$ [Fig.~\ref{fig:scaling}(a)], giving strong indication of KPZ superdiffusive wandering of domain walls between different specialists, which is characterized by $z=3/2$. \red{This scaling persists for small $\gamma$ [for $\gamma=0.1$, Fig.~\ref{fig:scaling}(a) yields $z=1.501\pm0.003$] before the crossover to possibly exponential takeover [Fig.~\ref{fig:scaling}(c) and next paragraph].} Higher-order cumulants of $\tau_f$ are also approximately multiples of $3/2$, further supporting superdiffusive behavior~\redd{\cite{Note5}}. These observations recapitulate the superdiffusive wandering of bacterial sectors observed experimentally in Ref.~\cite{hallatschek2007genetic}. 

For bounded specialization [Fig.~\ref{fig:regimes}(d)] and $\gamma=0$, we observe  $z=2.004\pm0.003$ for the scaling of $\langle\tau_f\rangle$ in $L$, strongly suggesting diffusive wandering of the boundaries between \red{specialized populations} (also supported by the scaling of higher-order cumulants of $\tau_f$ in~Ref.~\cite{Note5}). However, upon symmetry breaking with $\gamma\neq0$, diffusive behavior is no longer observed; rather, $\langle\tau_f\rangle$ appears to grow logarithmically with $L$ [Fig.~\ref{fig:scaling}(c)]. In particular, the inverse proportionality between $\gamma$ and the slope in the log-plot suggests that the size of the domains grows proportional to $e^{\gamma t}$. However, with unbounded fitness gain and $\gamma\neq0$, the results are too broadly distributed to draw definitive conclusions.

The transition between the universality classes of bounded and unbounded fitness is best illustrated in the change from Fig.~\ref{fig:regimes}(d) to Fig.~\ref{fig:regimes}(c) for $\alpha>0$ as $\beta$ changes sign. As depicted in the left panel of Fig.~\ref{fig:scaling}(b), the corresponding switch from $z=2$ to $z=3/2$ is quite abrupt.
On the other hand, the fitness function for $\alpha<0$ and $\beta>0$ in Fig.~\ref{fig:regimes}(b) leads to more complex evolutionary dynamics: an initial linear growth of emerging \red{specialists} in the unspecialized background followed by their competition towards final fixation. At later times, the competition between specialists with unbounded growth of fitness resembles the dynamics of Fig.~\ref{fig:regimes}(c). The effective exponent, depicted in the right panel of Fig.~\ref{fig:scaling}(b) reflects this two-step approach to fixation, with $z$ gradually nearing 3/2 as the second stage becomes more prominent upon the increase of $\beta$.

\textit{Discussion.} In summary, we investigate a model for \red{specialization} at the front of a colony expanding synchronously into new territory. By accumulating mutations, progenitors of the initial seed population \red{differentiate into distinct specialists;} the driving force for \red{specialization} is the gain or loss in fitness upon acquiring these mutations. With competition and selection occurring only locally on the growth front, the emerging \red{specialists} initially segregate into sectors, which subsequently expand to dominate the entire population. By quantifying the scaling of fixation time with population size, we find that unbounded and bounded gains in fitness lead to superdiffusive ($z=3/2$) and diffusive ($z=2$) stochastic wanderings of the sector boundaries, respectively; that is, an unbounded fitness gain in this setting leads to more rapid fixation, but with a distinct mathematical characteristic. It remains to show if this distinction is robust to variations of the model, such as with asynchronous growth.

While removed from reality, simplified models as pursued here point to relevant features and their importance; the emergence of ``complexity'' in such models is typically classified in terms of universality classes that share gross underlying features. In the context of \red{specialization}, future work to explore more complex fitness landscapes involving multiple traits may elucidate more complex evolutionary dynamics. The dimensionality of the space over which the colony expands, as well as environmental heterogeneities~\cite{mobius2015obstacles}
are also factors to consider. Additional effects to explore include changes in habitat ranges~\cite{nullmeier2013coalescent}, mutualistic or antagonistic interactions between \red{specialized populations}~\cite{lavrentovich2014asymmetric}, environments with curved surfaces~\cite{beller2018evolution}, and successive range expansions~\cite{goldschmidt2017successive}.

Ultimately, our model presents one possible mechanism for individuals to \red{diversify and specialize} in expanding populations. Our findings corroborate natural and experimental observations and has the potential to predict evolutionary phenomena occurring in systems where it is disadvantageous to specialize in multiple features. \red{Future investigations can explore a two-parental model to see if similar growth patterns emerge.} We hope our work will inspire further investigations into evolutionary dynamics on these frontiers.

\textit{Acknowledgments}. M.K.~is supported by NSF Grant No.~DMR-\red{2218849}. Simulations were performed using services provided by the OSG Consortium~\cite{ruth2007open,sfiligoi2009pilot}, which is supported by the NSF Awards No.~2030508 and No.~1836650.  Source code used in the preparation of this manuscript is in Ref.~\footnote{Source code can be found at \url{https://github.com/laurenhli/2022-spec}}.

\balance
\bibliography{papersources}


\clearpage

\setcounter{equation}{0}
\setcounter{figure}{0}
\setcounter{table}{0}
\setcounter{page}{1}
\makeatletter
\renewcommand{\theequation}{S\arabic{equation}}
\renewcommand{\thefigure}{S\arabic{figure}}
\renewcommand{\thepage}{S\arabic{page}} 
\renewcommand{\thesection}{S\arabic{section}} 
\renewcommand{\bibnumfmt}[1]{[S#1]}
\renewcommand{\citenumfont}[1]{S#1}

\title{Supplemental Materials: \red{Specialization} at an expanding front}

\maketitle
\onecolumngrid
\vspace{-45pt}
\section{Emergence time} \label{sec:emerge}

\begin{figure}[H]
    \centering
    \includegraphics{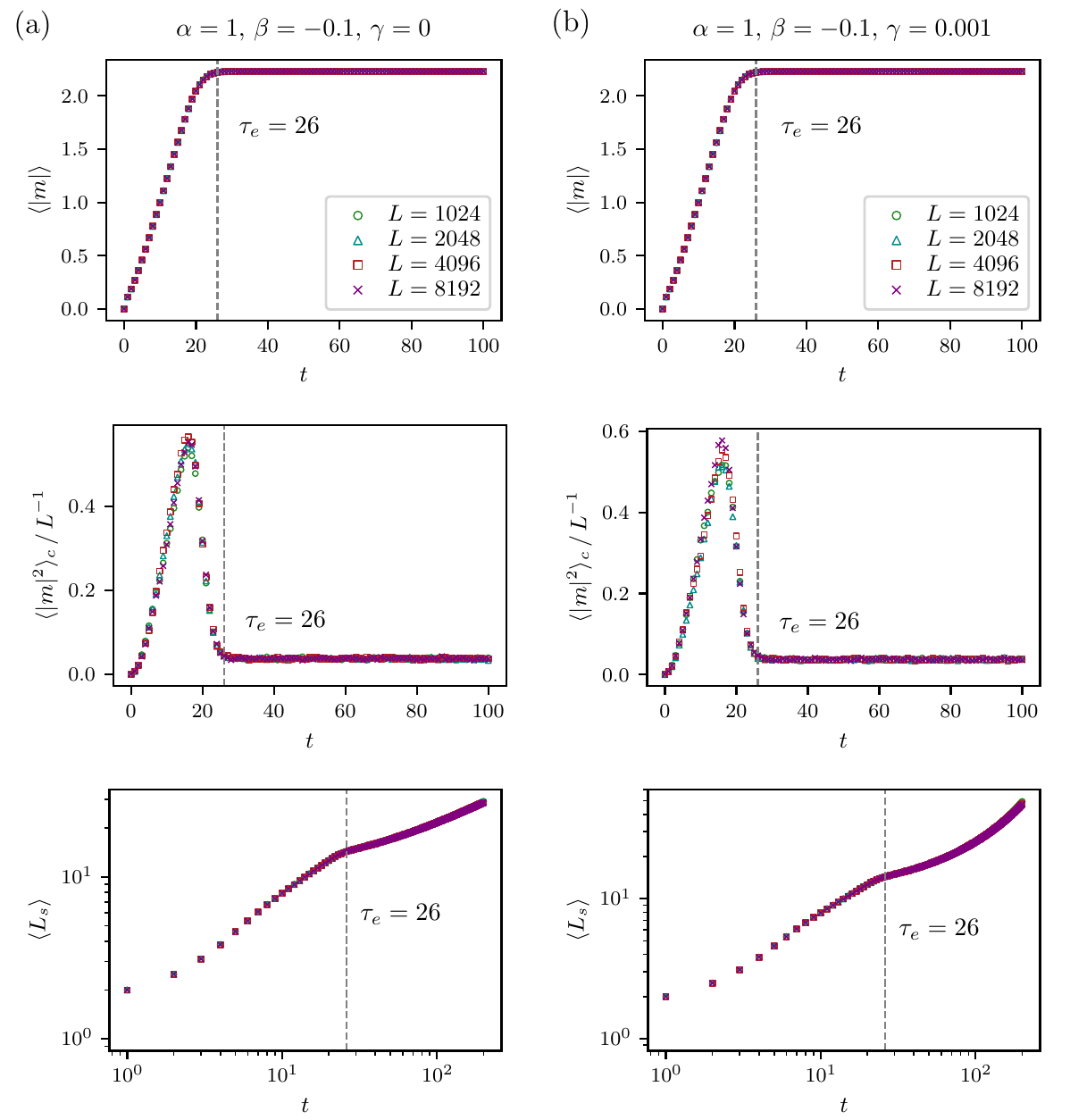}
    \caption{Emergence time in the bounded fitness model of Fig.~\ref{fig:regimes}(d). Similar trends are observed for (a) $\gamma=0$ and (b) $\gamma=0.001$. All plots demonstrate an emergence time $\tau_e=26$. We set the standard deviation of the mutational variability $\eta$ to $\sigma=0.1$ in all plots. Statistics for $\abs{m}$ and $\langle L_s\rangle$ are calculated over $10^3$ and $10^4$ realizations, respectively.}
    \label{fig:emerge}
\end{figure}

If the fitness gain is bounded as in Fig.~\ref{fig:regimes}(d), there is a typical emergence time scale $\tau_e$, during which the degree of specialization $\abs{m}$ grows and saturates. Fig.~\ref{fig:emerge} depicts features of this saturation process: there is a linear increase in $|m|$, which stops at a characteristic $\tau_e$. The variance of $|m|$ also comes to saturation, but, interestingly, is non-monotonic with a pronounced peak prior to specialization. We note that the statistics here are calculated over $L$ sites at a given generation, and the variance collapses when divided by $L^{-1}$.

We also define the characteristic size $\langle L_s\rangle$ of a specialized population at a given generation by $L/n_s$, where $n_s$ is the number of sector boundaries (except in the case with no sector boundaries, in which $n_s=0$ and $\langle L_s\rangle=L$). We observe that the characteristic size shows two different scaling behaviors in time (Fig.~\ref{fig:emerge}). In particular, we identify the transition at $\tau_e$ by the crossover point, with $\langle L_s\rangle$ increasing at different rates before and after this emergence time. Furthermore, the data collapse of $\langle L_s\rangle$ indicates that the crossover time is independent of system size, and that the size of specialist subpopulations also grows independent of system size until $L$ is reached. 

\section{Higher-order cumulants} \label{sec:cum}

\begin{figure}[H]
    \centering
    \includegraphics{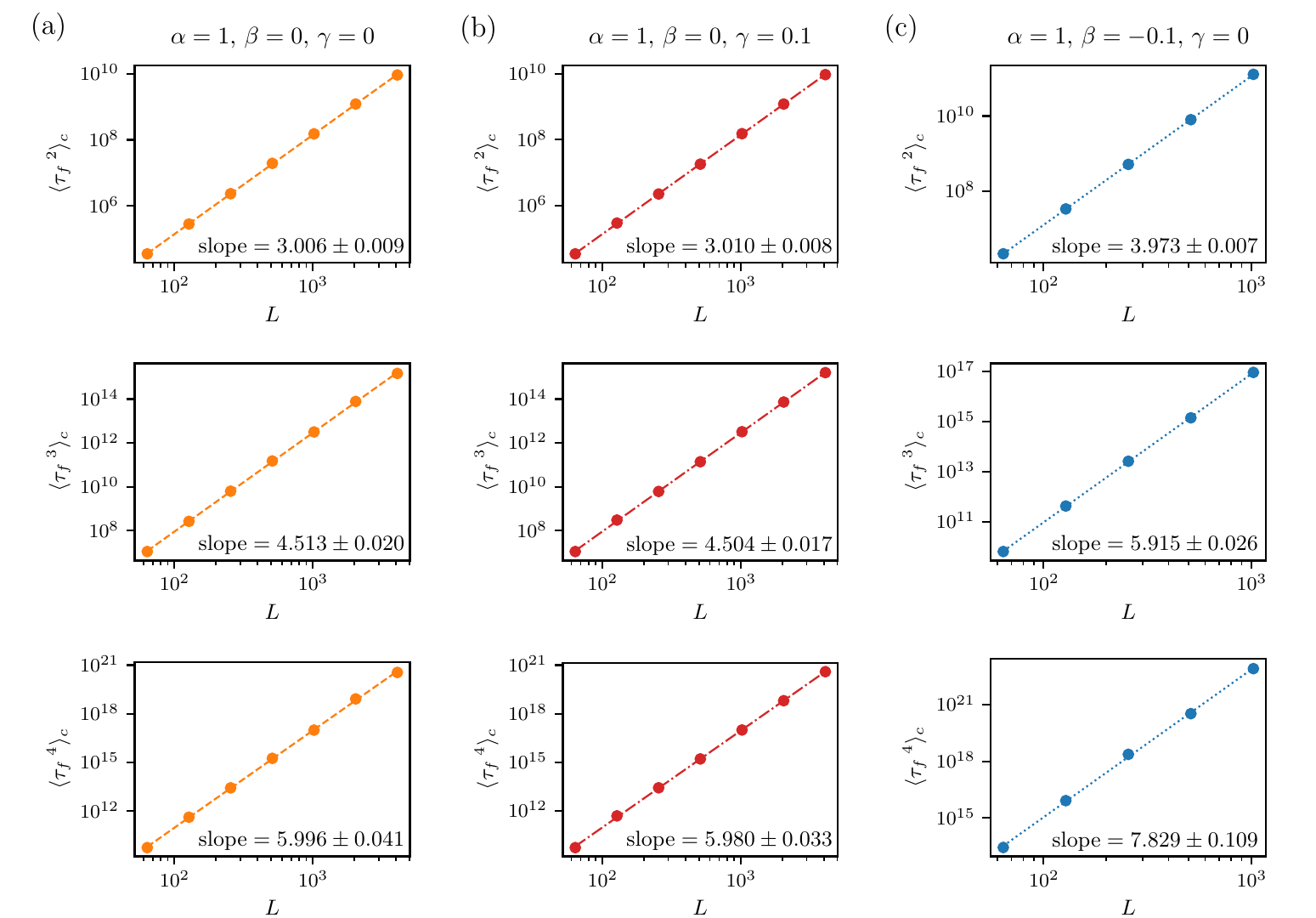}
    \caption{Log-log plots of higher-order cumulants of the fixation time with $L$. Slopes are roughly integer multiples of $z=3/2$ in (a) and (b); and of $z=2$ in (c). Statistics are calculated over $10^4$ realizations, with $\sigma=0.1$ in all plots.}
    \label{fig:cum}
\end{figure}

We observe that the distribution of fixation times $\tau_f$ across many realizations is not Gaussian. Rather, the higher-order cumulants also reflect the dynamical scaling exponent $z$; we find that \red{the slopes of} the higher-order cumulants are approximately integer multiples of $z$ (Fig.~\ref{fig:cum}).

\section{Fluctuations of sector boundaries}

\begin{figure}[H]
    \centering
    \includegraphics{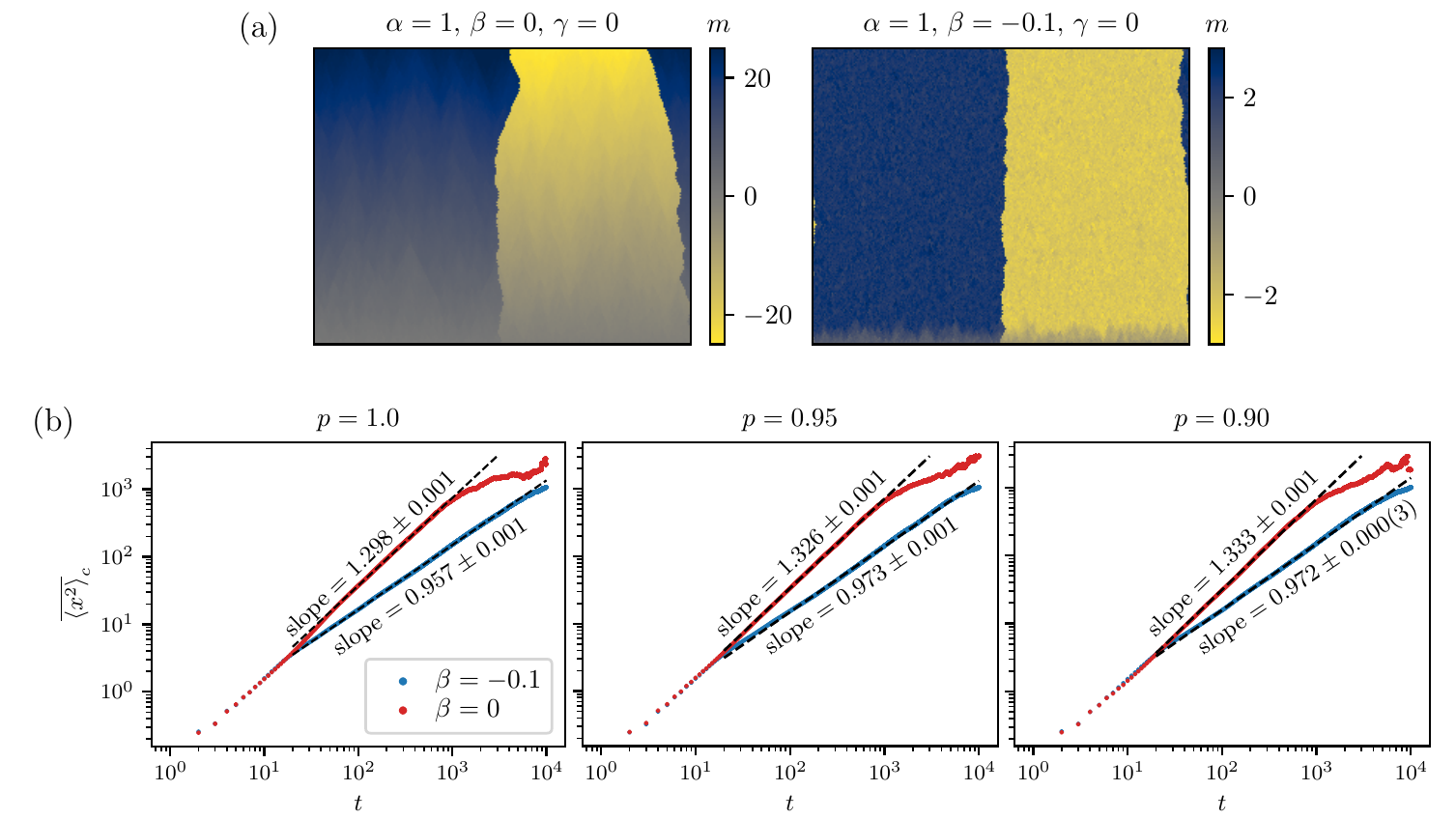}
    \caption{\red{Fluctuations of sector boundaries. (a) Half of the initial population is initialized to specialists in $b$ or $h$\redd{; labeling the leftmost site as $x=0$, we initialize $m(x,0)=1$ for $x < L/2$, and $m(x,0)=-1$ for $x\geq L/2$}. Color plots show $m(x,t)$ for populations of size $L=256$ over 200 generations with time running in the upwards direction. (b) Mean square transverse displacements of the sector boundaries over time; here, we set $\alpha=1$ and $\gamma=0$. Statistics are calculated over $10^3$ realizations, with $\sigma=0.1$ in all plots.}}
    \label{fig:half}
\end{figure}

\red{To show that the fixation time behaviors are indeed a reflection of the sector boundary wandering, we simulate the growth of two specialized populations, each occupying one half of the space [Fig.~\ref{fig:half}(a)]. We then quantify the mean square transverse displacements of the sector boundaries, and fit to scaling in time as
\[
\langle x^2\rangle_c \sim t^{2\zeta}.
\]
For unbounded fitness, we find $\zeta=0.649\pm0.001\approx2/3$, consistent with KPZ superdiffusive behavior; for bounded fitness, we find $\zeta=0.479\pm0.001\approx1/2$, which is characteristic of diffusive wandering [Fig.~\ref{fig:half}(b)]. As the coalescence of wandering domains is the mechanism for fixation, the above result provides further support for the use of fixation time to infer the universality class. }

\red{In the main text, we only consider randomness introduced \redd{by mutations} during the evolution of $b$ and $h$. However,  randomness may also \redd{occur due to environmental fluctuations and affect the selection of} the site that propagates. Thus, given two competing neighboring cells, we consider the probability $p$ that the cell with the larger fitness $f(b,h)$ will reproduce; in the case of $p=1.0$, we recover the same model as in our main text. For values of $p<1$, we see that superdiffusive and diffusive behaviors are still observed with unbounded and bounded fitness, respectively [Fig.~\ref{fig:half}(b)].} \redd{We note that there are other potential sources of noise, and we leave investigations of their effects to the future.}

\end{document}